\documentclass[a4paper,10pt]{article}

\usepackage{hyperref}
\usepackage{breakurl}
\usepackage{indentfirst}
\usepackage{authblk}

\title{twister - a P2P microblogging platform}

\author{Miguel Freitas%
  \thanks{e-mail: \texttt{miguel@cpti.cetuc.puc-rio.br}}}
\affil{Center for Research in Inspection Tecnology (CPTI)\\
        Pontifical Catholic University of Rio de Janeiro (PUC-Rio), Brazil}

\begin{document}

\maketitle

\begin{abstract}

This paper proposes a new microblogging architecture based on peer-to-peer networks overlays. The proposed platform is comprised of three mostly independent overlay networks. The first provides distributed user registration and authentication and is based on the Bitcoin protocol. The second one is a Distributed Hash Table (DHT) overlay network providing key/value storage for user resources and tracker location for the third network. The last network is a collection of possibly disjoint ``swarms'' of followers, based on the Bittorrent protocol, which can be used for efficient near-instant notification delivery to many users. By leveraging from existing and proven technologies, twister provides a new microblogging platform offering security, scalability and privacy features. A mechanism provides incentive for entities that contribute processing time to run the user registration network, rewarding such entities with the privilege of sending a single unsolicited (``promoted'') message to the entire network. The number of unsolicited messages per day is defined in order to not upset users.

\end{abstract}

\section{Introduction}

Microblogging platforms are one of the most versatile and empowering technologies on the internet today. Recent events have shown the important role of these tools for news coverage~\cite{sklar2009} and also for political movements, like in Middle East's ``Arab Spring''. Although their role in social revolutions should not be overstated~\cite{khondker2011role}, it is exemplary to learn how dictatorships frequently resort to shutting down the internet in trying to control such possibly destabilizing movements~\cite{glanz2011,warner2012}. Blocking internet access, however, is never fully effective against social movements, as some people always find ways to circumvent such blockage~\cite{dachis2011}.

The possibility that the service providers themselves could be convinced to participate of a social media blockage~\cite{halliday2011} would affect people's ability to communicate in a much more dramatic way than just disrupting a few network backbones. As our society's dependence on these services increases, the single point of failure on such basic communication platforms (at the provider's own discretion) is not only unacceptable but also directly opposes Internet's key design features of providing redundancy for information transmission~\cite{wikipediainternet}.

Reports of widescale internet wiretapping with the cooperation of large corporations~\cite{greenwald2013} reveal the danger the present platforms pose to user's privacy. The fact that a single entity is able to access private communication and personal data at their will should raise concerns to anyone who thinks about it for a while. A recent House of Lords (UK) report openly recognizes the dangers of mass surveillance\footnote{``Mass surveillance has the potential to erode privacy. As privacy is an essential pre-requisite to the exercise of individual freedom, its erosion weakens the constitutional foundations on which democracy and good governance have traditionally been based in this country.''\cite{surveillance}}.

All these facts point to an obvious direction: there is an urgent need for open, secure and distributed personal communication platforms. This is where the present peer-to-peer microblogging proposal fits in.

Of course, to be sucessful, such P2P microblogging cannot just provide resilience and security, but it must also be user-friendly. This is a key point to the adoption of any new software or web service. Some current P2P message proposals offer good examples of what not to do in terms of user-friendliness, like requiring the user to know a cryptic address composed of 36 case sensitive characters~\cite{warren2012bitmessage}.

The ability to provide easy to remember user logins must be considered a fundamental requirement. While users must be free to choose their login names, providing anonymity to whoever needs to express himself freely without fear of retaliation, it is important to realize that a web of trust is built on these microblogging infrastructure based upon real existing and fully identifiable people. This issue can be appreciated in Hudson plane crash coverage~\cite{sklar2009} where trusted aggregators helped separating the reliable information from random noise. These people tend to work as hubs in information circulation and are often defined as ``influential''. Any serious P2P microblogging proposal must foster this kind of organization.

This paper presents a proposal of a new P2P microblogging platform that is scalable, resilience to failures and attacks, does not depend on any central authority for user registration, provides easy-to-use encrypted private communication and authenticated public posts. The architecture tries to leverage from existing and proven P2P technologies such as Bittorrent and Bitcoin as much possible. Privacy is also one of the primary design concerns, no one should be able to see the user's IP or their followers unless he explicitly shares such information.

The proposed platform is comprised of three mostly independent overlay networks. The first provides distributed user registration and authentication and is based on the Bitcoin protocol. The second one is a Distributed Hash Table (DHT) overlay network providing key/value storage for user resources and tracker location for the third network. The last network is a collection of possibly disjoint ``swarms'' of followers, based on the Bittorrent protocol, which can be used for efficient near-instant notification delivery to many users.

\section{Related work}

Existing social networks like Diaspora, StatusNet and identi.ca are frequently cited as free, distributed alternatives to Facebook or Twitter. These platforms are based on the concept of ``federated social networks''~\cite{wikipedia_distributed} where users may join the social websites of their choice and these sites communicate with each other using open protocols. While technically superior to a single, closed platform in terms of achieving better privacy control, the user still needs to delegate his own data to a third party (unless he wants to setup his own server to federate).

Previous P2P microblogging proposals exist like Cucko~\cite{xu2010cuckoo} and Megaphone~\cite{perfitt2010megaphone}. Neither Cucko or Megaphone seems to address the problem of decentralized user registration. Privacy is also not one of Cucko's objectives since it is explicitly designed to know about the online presence of anyone. One similarity of twister and Cucko is that both share the idea of using an unstructured overlay network for dissemination of user posts, unlike Megaphone where all followers must register to the sender, forming a multicast tree for post propagation.

A more advanced social network proposal Safebook~\cite{cutillo2009safebook} address several privacy issues by implementing different levels (``shells'') of access to the published data. While Safebook's scope goes far beyond twister's, it still relies in a centralized Trusted Identification Service for user registration.

At the present time, no public implementation seems to be available to any of these P2P proposals.

\section{Notation}

Tuples (concatenation): $ [a,b,c,...] $

Apply function $f$ to payload $x$: $ y = f(x) $

Address of user $j$: $ ID_j = H(Username_j) $; where $H$ is hash function.

Keys of user $j$: $ PUBK_j; PRIVK_j $

Note: $PUBK(PRIVK(x)) = x$ and $PRIVK(PUBK(y)) = y$

Signed content $x$ from user $j$: $SIG_j(x) = \left[ PRIVK_j( H(x) ), x \right]$

\section{User registration P2P network}

Decentralized yet secure user registration is achieved by means of the Block Chain mechanism, which is used in Bitcoin~\cite{nakamoto2008bitcoin} to avoid the double-spending problem without the need for a central authority. In the proposed system the mechanism is used to guarantee the uniqueness of users, again with no need for a central authority. New registrations must be ``notarized'' by a number of Blocks before they can be considered granted to a given user. Each Block is defined as:

$Block_i = \left[ i, H(Block_{i-1}), Nonce_i, SpamMsg_i, \left[UserReg_j, UserReg_{j+1}, ...\right] \right]$

$H(Block_i)$ produces Proof-Of-Work (POW) due to partial hash collision (by brute force searching over $Nonce_i$ space). Difficulty is automatically set by the network based on the average number of blocks per hour (same as Bitcoin).

$UserReg_j = [Username_j, PUBK_j, Nonce_j]$

A new user $j$ registering to the network must broadcast $UserReg_j$. Other Nodes, upon receiving $UserReg_j$, must check POW as a partial hash collision of $H( UserReg_j )$ before the request can be retransmited/accepted. This POW prevents DoS attacks due to flooding of bogus registrations. POW of $UserReg_j$ is much smaller than POW of block chain, tipically just a few minutes of an average computer time (difficulty may be hardcoded in software and change only with protocol versions).

Block chain provides public dictionary from $Username_j$ (or $ID_j$) to $PUBK_j$.

Nodes must enforce the uniqueness of $Username_j$ before incluing $UserReg_j$ into a new Block. The only exception to this rule is the key replacement case, where the new public key is signed by the previously known key pair. The enforcement of uniqueness of $ID_j$ and POW of $UserReg_j$ is also applied when receiving new Blocks, since all registrations included therein must be checked.

$Username_j$ is also subject to additional text rules, such as maximum size and allowed characters. This further protects $ID$ space from partial hash collision which might, otherwise, allow monitoring (wiretapping, see section~\ref{sec:remarks}).

$SpamMsg_i$ is an unsolicited message (commonly and euphemistically called ``promoted'') that must be shown by all clients and provides incentive for joining the Block generation effort. If the same Bitcoin's block creation rate is maintained (6 per hour), a display probability factor may be implementated in order to not upset the users with too much spam. 

Developers must not implement hiding of spam messages as a ``feature'' of their clients since this incentive is important to the security of the entire network. Omitting unsolicited messages from clients would only hurt the users in the long run. A display probability factor may implemented, however, and the client might priorize localization (by giving higher probability to messages of the same language of the user) to improve effectiveness and also user's experience.

\section{Routable DHT network overlay}
\label{sec:dhtnet}

The second P2P network is a structured Distributed Hash Table (DHT) overlay network like Kademlia~\cite{maymounkov2002kademlia}. The single most important feature of this network is to allow resource storage and retrieval by peers. Direct delivery of notification between users can be thought as a secondary usage (see section~\ref{sec:mentioning}).

It would be tempting to use $ID_j$ directly as the address of the peer joining the DHT network, as it would permit simple challenge-response authentication, possibly preventing $ID$ forgery. Forged $ID$ address is arguably the most serious security issue on P2P/DHT networks (see Sybil and Eclipse attacks~\cite{wang2006attacks,yang2012survey}). Using $ID_j$ for DHT addressing, however, would greatly compromise privacy since it is a fundamental characteristic of such a network to know the $IDs$ of the other nodes in order to create optimized routing tables. $ID_j$ not only would allow easily detection of online user presence but would also reveal his IP address.

Instead of $ID_j$, the proposal is to use the standard procedure of hashing IP address and port number to join the DHT network:

$ID_{node\_j} = H\left( [IP_j,port] \right)$

In~\cite{dinger2006defending} it is shown that a secure mapping of external IPs to $ID$ is Sybil-proof when limited per participant.

Packets on this DHT network sent from $ID_{src}$ to $ID_{dst}$ are defined as follows:

$Packet = \left[ ID_{dst}, ID_{src}, SIG_j( payload ), ID_j) \right]$

The payload is signed by a given user $ID_j$, even though it may differ from the sender $ID_{src}$ in case of packet being retransmited/refreshed. These characteristics comprise the basic ``layer 3'' functionality offered by this overlay network.

Going up in the conceptual model for the proposed DHT overlay network there is an ``application layer'' with a data storage primitive (PUT) defined with the following payload:

$payload_{PUT} = [ target, value, time, seq ]$ where

$target = [owner, resource, restype]$ and $ID_{dst} = H(target)$

Some simple rules must be checked by destination node in order to accept the storage request:

\begin{enumerate}
    \item $ID_{dst} = H(target)$ : ensures the destination address was properly computed.
    \item $ID_{dst}$ is neighbor of $ID_{node}$ that actually received this request (by some agreed metric).
    \item $ID_j = H(owner)$, only enforced for $restype = $``$single$''.
    \item $seq$ is greater previously stored $seq_{old}$, only enforced for $restype = $``$single$''.
    \item $time$ is a valid time (ie, not in future).
\end{enumerate}

The two possible $restype$ values are ``single'' and ``multi''. These two types provide, respectivelly, resources which may only be updated by the owner of this key (like an avatar image) and resources which collect multiple responses from different users (like replies to a certain post). In case of the ``single'' type, the node stores just a single $value$ associated with this key $ID_{dst}$. For ``multi'', however, new PUT requests are appended to a list of $value$'s. This kind of storage provides no guarantees, values may be discarded following expiration (based on $time$ field) or Least Recently Used (LRU) cache strategy. Authenticated (``single'') storage takes precedence over any previously ``multi'' value.

A data retrieval primitive (GET) may operate on both types of resources indistinctively. Some special non-storage resources associated with dynamic content may also be implemented using the same primitives, thus sharing the same API.

\section{User posts}
\label{sec:userposts}

The $k$-th message of user $j$ is defined as:

$UserPost_{jk} = SIG_j( [Username_j, k, type, MSG_k, REPLY_k] )$

where $MSG_k$ is the content (140 characters limited), $k$ is a monotonic increasing number and $type$ may define if it is a new post, a reply, retransmission (RT) or Direct Message (DM). $REPLY_k$ is an optional field which provides a reference pointer to the original message, in case of a reply/RT (see section~\ref{sec:navigate_top}) and is defined as tuple $REPLY_k = [Username_{j'},{k'}]$, where original post is the $k'$-th message of the user $j'$.

The posts are shared simultaneously in two overlay networks: (1) as a stored value, possibly short lived, in DHT network and (2) in a file-like archive pertaining to a kind of Bittorrent network. When a new post is created, the client must send two PUT requests to the following addresses:

$ID_{UserPost\_jk} = H\left( [Username_j, \textnormal{``post''} + k, \textnormal{``single''}] \right)$ and

$ID_{swarm\_j} = H\left( [Username_j,\textnormal{``swarm''}, \textnormal{``single''}] \right)$. 

The $ID_{UserPost\_jk}$ is the address of a storage target defined in section~\ref{sec:dhtnet} and provides arbitrary post retrieval capabilities.

The $ID_{swarm\_j}$ is a special gateway address to reach a torrent swarm (in Bittorrent nomenclature~\cite{bittorrentterms}). This torrent may contain all posts from a given user $j$ and helps sharing them independently of the DHT network. The neighbors of $ID_{swarm\_j}$ are required to join this swarm, as much as the neighbors of $ID_{UserPost\_jk}$ are required to store the value. The DHT-torrent interaction rules are further detailed in section~\ref{sec:swarm}. 

The swarm mechanism for distributing new posts fixes the problem of efficient notification of new posts, sparing the followers the need to do polling on a certain address of the DHT network. This is a different solution for the same issue (``lame, repeated polling'') raised by developers behind the pubsubhubbub protocol~\cite{pubsubhubbub}.

\subsection{Direct Messages}

User posts may also be used to send Direct Messages (DM), provided that recipient is a follower of user $k$ (same requirement as Twitter).

$UserPost(j \rightarrow l)_k = SIG_j\left( \left[\textnormal{``''}, k, \textnormal{``dm''}, [ PUBK_l(DM_k), H(DM_k) ] \right] \right)$

One should note that DM is equivalent to a normal post except that 

$[PUBK_l(DM_k), H(DM_k)]$ replaces the usual public message payload above.

DM is only received by destination user $l$ by checking for sucessful decryption. No other user will know for which recipient the DM was sent to, although the encrypted message will be seen by all his followers.

This naive description of DM encryption mechanism is only meant to explain the concept and the actual implementation may differ. Currently, the working twister prototype is based on an ECIS (Elliptic Curve Integrated Encryption Scheme) implementation by Ladar Levison~\cite{levison2010} (formerly the owner of Lavabit encrypted email service) supposedly following the SECG SEC1 standard~\cite{secg2009sec1}.

\subsection{User posts torrent/tracker rules}
\label{sec:swarm}

\begin{itemize}
 \item Online neighbors of $ID_{swarm\_j}$ within a certain distance in hash space are required to join (or create) the swarm.
 \item When a neighbor of $ID_{swarm\_j}$ receives a new post from the DHT network he must act like a gateway, incorporating the posts into the file-like structure shared by the Bittorrent network.
 \item The Bittorrent tracker is a special ``read-only'' multi-value list storage addressable by
          $ID_{tracker\_j} = H\left( [Username_j,\textnormal{``tracker''}, \textnormal{``multi''}] \right)$
 \item Followers of user $j$ should join the swarm to receive real-time updates. To do so they query $ID_{tracker\_j}$ (GET primitive) for a list of initial peers.
 \item The $ID_{tracker\_j}$ differs from other storage keys because it's read-only nature. This is a security measure to prevent tracker poisoning and also to protect privacy of swarm members. The list of IP addresses is therefore obtained from the swarm protocol itself (Bittorrent) instead of being writtable from the DHT network. This adds an additional requirement though: online neighbors of $ID_{tracker\_j}$ are required to join the swarm as well. 
 \item Swarm members only know each other by their IP addresses. This Bittorrent-like network must provide no hint of their $Usernames$.
 \item A table of hashes of all user posts (ie. like Torrent's pieces checksums) is not needed since all posts (including DMs) are already signed and can be verified.
 \item Increment in $k$ (new post) is be propagated directly by broadcast within the swarm (flooding).
 \item Swarm members exchange bitlists of available posts. Members may choose to only keep/request the most recent posts.
 \item Seeders are nodes who choose to be archivists.
 \item The producer (the user $j$) may choose not to be member of his own swarm (for privacy purposes, protecting his IP).
 \item If the producer chooses to be member of the swarm, he might skip entirely the $ID_{swarm\_j}$ gateway scheme, losing some IP privacy.
 \item Even if a producer is a member of the swarm, he does not need to be a seeder.
 \item In Bittorrent terminology, the number of existing pieces must be increase to $k$ with new posts. This is achieved by sending an (unsolicited) ``have'' message.
 \item Clients must regard the parameter of a ``have'' message as the new number of pieces. In order to prevent Denial-Of-Service attacks, this number is constrained by $k < 2 * (iBlock_{current} - iBlock_{User\_reg}) + 20$. The number is recused otherwise.
 \item If a new Block $k$ is produced every 10 minutes this limits the mean post rate of new users, for life, to a maximum of 288 posts/day. Average.
\end{itemize}

\section{Mentioning}
\label{sec:mentioning}

If a message mentions the user $j$ in a new post (@username) the client must also send a notification to $ID_j$, by including the full message. Notification is routed by DHT network.

Mentioning is the only feature in the proposed architecture which would require to route packets to a specific user addressed $ID_j$, not $ID_{node_j}$. Alternatively, a different 

$ID_{mention\_j} = H\left( [Username_j,\textnormal{``mention''}] \right)$

could be set to receive and accumulate mentions, to be maintained by nodes which are neighbors of $ID_{mention_j}$. The only issue here is again the ``lame, repeated polling'' as the user would need to periodically pool such key (although in a much more limited scale than a hashtag, for example).

A way to prevent pooling of user mentions while preserving some degree of privacy is to elect ``listeners'' for an $ID_j$ destination. Those listeners would then forward the packets to the final user. The idea is partially based in SASON~\cite{tsai2006scalable}, although not as secure since an additional anonymizing network is not used.

The system would work like this: the recipient $ID_j$ first uses the DHT network to find nodes near $ID_j$. He then asks them directly to forward all $ID_j$ traffic to $ID_{node\_j}$, therefore revealing his real identity to a small group of listeners. Listeners must do a challenge validation to make sure the user is really $ID_j$ by asking for $SIG_j( \textnormal{random number} )$. Since the other node has access to the full directory of public keys, he can easily authenticate.

Mentioning, like other mechanisms described here, requires the cooperation of the client software in order to work. If a given user does not send the notification packet to the network (along with his own post) the mentioned user would never know.

\section{Explict message request}

User $l$ may request explicitly a certain message from user $j$ without joining the swarm. This is achieved by a simple authenticated value retrieval from address $ID_{UserPost-jk}$.

This feature allows for ``upward message thread navigation'' like in Twitter and is not resource intensive.\label{sec:navigate_top}

\section{Downward message thread navigation}

Downward navigation (finding out about replies/RT of a certain post) might be a difficult problem since there are many, possibly unlimited, answers to the question ``what are the replies to this particular post?''.

One possible solution is to send another notification to the special address of multi-value list storage:

$ID_{replies\_jk} = H\left( [Username_j, \textnormal{``replies''}+k, \textnormal{``multi''}] \right)$

The values to store are copies of the replies themselves ($UserPost$ format defined in section~\ref{sec:userposts}). Again, it is the cooperation of client posting the reply which allows this mechanism to work.

\section{Hashtags}

Just like mentioning, hashtags must be detected in the content of new messages being posted to the network. A copy of the message is sent to a special address of multi-value list storage:

$ID_{hashtag_t} = H\left( [hashtag_t, \textnormal{``hashtag''},  \textnormal{``multi''}] \right)$

This is pretty much the same mechanism of downward message thread navigation except for an additional feature: a hashtag creates a new swarm similar to  $ID_{swarm\_j}$. Neighbors of such $ID_{hashtag_t}$ are be forced to join this virtual swarm which has no sequential content (file). Posts that include the hashtag are DHT routed to a neighbor member of the swarm, from which they are broadcast to the swarm's members.

This swarm is therefore just used to create a distributed tracker and broadcast mechanism for users willing to monitor such hashtags. New members joining the swarm may also request the last messages from the multi-valued storage (DHT network), without guarantee of completeness.


\section{Word search}

Searching for arbitrary words may be achieved by extending the hashtag implementation concept to all words in every post. In order to reduce overhead and network traffic, certains limits can be imposed like minimum word size, excluding prepositions and so forth. 

Another difference to hashtags is that creating swarms for all possible words may be considered an overkill. So the collection of posts containing a given word would be limited to a temporary multi-value list storage addressable by

$ID_{word_w} = H\left( [word_w, \textnormal{``word''},  \textnormal{``multi''}] \right)$

\section{Concluding remarks}
\label{sec:remarks}

The proposed architecture provides a distributed P2P microblogging network with security, scalability and privacy features.

\begin{itemize}
 \item The architecture is resilient like other P2P technologies, so it is believed that no single company, government or other entity should able shut it down.

 \item The distributed user registration mechanism is secure like Bitcoin transactions, providing content authentication without relying on any particular entity.

 \item Real existing people have an incentive for early adoption in order to choose their username of preference.

 \item Using common usernames, instead of long cryptographic hashes as seen in some other proposals, makes the system as user-friendly as existing microblogging systems.

 \item Public key replacement allows one to change one's own key pair when security is compromised in any way (eg. stolen cell phone). It also makes possible for users or companies to buy their usernames of choice (like existing domain names commerce).

 \item Main features of other existing microblogging systems are replicated, including simple username search, thread navigation, mentioning, private messages, hashtags and word search.

 \item The DHT routing provides a way to send notifications to, and request resource data (avatar image, profile etc) from a particular user, without knowing if he is online or not.

 \item In order to detect a user's IP address or interfere/spy his activities, an entity would need to try assigning itself an $ID_{node}$ which is close to the victim's (or one of his associated resources, like trackers). Because restrictions imposed on $ID_{node}$ validation from external IP addresses, this is not an easy task. 

 \item An entity in possession of big resources (lots of blocks of IPs to choose from) might be able to achieve this partial $ID_{node}$ collision to spy on the activities from of a specific user. This moves the wiretapping capabilities from ``mass surveillance'' to the much more reasonable ``targeted surveillance'' (see~\cite{surveillance} for definitions).

 \item While finding an user's online presence might be difficult, this is not a strict guarantee of this architecture. Users demanding further privacy are suggested to use twister on top of Tor~\cite{torproject}.

 \item The architecture provides incentive for enterprises to run the system infrastructure in order to have the right to send promoted messages at a very limited rate. While this may be commerced as advertisement, it also allows groups of users to join in a community effort to try to spread some information out (like Bitcoin mining pool). So the proposal is also quite democratic.

\item Independent providers may offer twister access from a standard web interface, joining the P2P network on the backend. However, while perfectly legal and supported, this model defeats most of the privacy and security features since the provider would be in possession of the user's $PRIVK$.

\item A clever evolution of the web interface to twister network would store a password-encrypted version of $PRIVK$ in the server so that, in order to send new messages, the key is temporary decrypted by the javascript running within the browser. This idea would prevent the server's owner from being able to impersonate the user.

\item Read-only web interfaces for reading users' public posts and hashtags are possible and do not compromise security.

\item Resource-limited clients (like mobile phones) may choose to work with some optimizations. For example, they may not store the full block chain but rather just the chain of block hashes. In order to search for a particular user they might ask the network which block specifically contains this user registration. Then the client would download just the required block without incurring any loss of security (the block integrity is verifiable). Instead of a full block download a partial Merkle tree can be used.

\end{itemize}

\bibliographystyle{unsrt}
\bibliography{twister}

\end{document}